\documentclass{elsart}
\usepackage{graphicx}
\usepackage{amsmath}
\usepackage{amsfonts}
\usepackage{amssymb}
\begin{document}
\begin{frontmatter}
\title
{Irreducible mass and energetics of an electromagnetic black hole}
\author[icra]{Remo Ruffini}\ead{ruffini@icra.it},
\author[icra]{Luca Vitagliano}\ead{vitagliano@icra.it}
\address
[icra]{International Centre for Relativistic Astrophysics, Department of Physics,
Rome University ``La Sapienza", P.le Aldo Moro 5, 00185 Rome, Italy}
\begin{abstract}
The mass-energy formula for a black hole endowed with electromagnetic structure (EMBH) is clarified for the nonrotating case. The irreducible mass $M_{\mathrm
{irr}}$ is found to be independent of the electromagnetic field and explicitly expressable as
a function of the rest mass, the gravitational energy and the kinetic energy of the collapsing matter at the horizon.
The electromagnetic energy is distributed throughout the entire region extending from the horizon of the EMBH to infinity. We discuss two
conceptually different mechanisms of energy extraction occurring respectively in an EMBH with electromagnetic fields smaller and larger than the critical field for vacuum polarization. For a subcritical EMBH the energy extraction mechanism involves
a sequence of discrete elementary processes implying the decay of a particle into two oppositely charged particles. For a supercritical EMBH an alternative mechanism is at work involving an electron-positron plasma created by vacuum polarization. The energetics of these mechanisms as well as the definition of the spatial regions in which thay can occur are given. The physical implementations of these ideas are outlined for ultrahigh energy cosmic rays (UHECR)
and gamma ray bursts (GRBs).
\end{abstract}
\end{frontmatter}

The main objective of this article is to clarify the interpretation of the
mass-energy formula \cite{CR71} for a black hole endowed with electromagnetic
structure (EMBH). For simplicity we study the case of a nonrotating EMBH using
the results presented in a previous letter \cite{CRV02}. The collapse of a
nonrotating charged shell can be described \cite{CRV02} by an exact analytic
solution of the Einstein-Maxwell equations. The world surface $S$ of the shell
divides the space-time into two complementary regions: an internal one
$\mathcal{M}_{-}$ and an external one $\mathcal{M}_{+}$. In spherical
coordinates the line element is
\begin{equation}
ds^{2}=\left\{
\begin{array}
[c]{r}%
-f_{+}dt_{+}^{2}+f_{+}^{-1}dr^{2}+r^{2}d\Omega^{2}\qquad\text{in }%
\mathcal{M}_{+},\ r>R\\
-dt_{-}^{2}+dr^{2}+r^{2}d\Omega^{2}\qquad\text{in }\mathcal{M}_{-},\ r<R
\end{array}
\right.  ,
\end{equation}
where $f_{+}=1-\tfrac{2M}{r}+\tfrac{Q^{2}}{r^{2}}$, and $t_{-}$ and $t_{+}$
are the Schwarzschild-like time coordinates in $\mathcal{M}_{-}$ and
$\mathcal{M}_{+}$ respectively. Here $Q$ is the charge of the shell and $M$
its mass-energy, measured by an observer at rest at infinity, while $R$ is the
coordinate radius separating the two regions and may be considered as a
function of either $t_{-}$ or $t_{+}$. $\mathcal{M}_{-}$ and $\mathcal{M}_{+}$
are static space-times; we denote their time-like Killing vectors by $\xi
_{-}^{\mu}$ and $\xi_{+}^{\mu}$ respectively. $\mathcal{M}_{+}$ is foliated by
the family $\left\{  \Sigma_{t}^{+}:t_{+}=t\right\}  $ of space-like
hypersurfaces of constant $t_{+}$.

The splitting of the space-time into two regions $\mathcal{M}_{-}$ and
$\mathcal{M}_{+}$ allows two physically equivalent descriptions of the
collapse and the use of one or the other depends on the question one is
studying. The use of $\mathcal{M}_{-}$ proves helpful for the identification
of the physical constituents of the irreducible mass while $\mathcal{M}_{+}$
is needed to describe the energy extraction process from the electromagnetic
black hole (EMBH). The equation of motion for the shell \cite{CRV02} is
\begin{equation}
\left(  M_{0}\tfrac{dR}{d\tau}\right)  ^{2}=\left(  M+\tfrac{M_{0}^{2}}%
{2R}-\tfrac{Q^{2}}{2R}\right)  ^{2}-M_{0}^{2} \label{EQA}%
\end{equation}
in $\mathcal{M}_{-}$ and
\begin{equation}
\left(  M_{0}\tfrac{dR}{d\tau}\right)  ^{2}=\left(  M-\tfrac{M_{0}^{2}}%
{2R}-\tfrac{Q^{2}}{2R}\right)  ^{2}-M_{0}^{2}f_{+} \label{EQAb}%
\end{equation}
in $\mathcal{M}_{+}$. $M_{0}$ is the total rest mass of the shell, $R$ is its
Schwarzschild radius and $\tau$ is the proper time along $S$. As remarked in
\cite{CRV02}, from the $G_{tr}$ Einstein equation we have the constraint
\begin{equation}
M-\tfrac{Q^{2}}{2R}>0. \label{EQO}%
\end{equation}
Since $\mathcal{M}_{-}$ is a flat space-time we can interpret $-\tfrac
{M_{0}^{2}}{2R}$ in (\ref{EQA}) as the gravitational binding energy of the
system. $\tfrac{Q^{2}}{2R}$ is its electromagnetic energy. Then equations
(\ref{EQA}), (\ref{EQAb}) differ by the gravitational and electromagnetic
self-energy terms from the corresponding equations of motion of a test particle.

Introducing the total radial momentum $P\equiv M_{0}u^{r}=M_{0}\tfrac
{dR}{d\tau}$ of the shell, we can express the kinetic energy of the shell as
measured by static observers in $\mathcal{M}_{-}$ as $T\equiv-M_{0}u_{\mu}%
\xi_{-}^{\mu}-M_{0}=\sqrt{P^{2}+M_{0}^{2}}-M_{0}$. Then from equation
(\ref{EQA}) we have
\begin{equation}
M=-\tfrac{M_{0}^{2}}{2R}+\tfrac{Q^{2}}{2R}+\sqrt{P^{2}+M_{0}^{2}}%
=M_{0}+T-\tfrac{M_{0}^{2}}{2R}+\tfrac{Q^{2}}{2R}. \label{EQC}%
\end{equation}
where we choose the positive root solution due to the constraint (\ref{EQO}).
Eq.~(\ref{EQC}) is the \emph{mass formula} of the shell, which depends on the
time-dependent radial coordinate $R$ and kinetic energy $T$. If $M\geq Q$, an
EMBH is formed and we have
\begin{equation}
M=M_{0}+T_{+}-\tfrac{M_{0}^{2}}{2r_{+}}+\tfrac{Q^{2}}{2r_{+}}\,, \label{EQL}%
\end{equation}
where $T_{+}\equiv T\left(  r_{+}\right)  $ and $r_{+}=M+\sqrt{M^{2}-Q^{2}}$
is the radius of external horizon of the EMBH. We know from the
Christodoulou-Ruffini EMBH mass formula that
\begin{equation}
M=M_{\mathrm{ir}}+\tfrac{Q^{2}}{2r_{+}}, \label{irrmass}%
\end{equation}
so it follows that {}%
\begin{equation}
M_{\mathrm{ir}}=M_{0}-\tfrac{M_{0}^{2}}{2r_{+}}+T_{+}, \label{EQM}%
\end{equation}
namely that $M_{\mathrm{ir}}$ is the sum of only three contributions: the rest
mass $M_{0}$, the gravitational potential energy and the kinetic energy of the
rest mass evaluated at the horizon. $M_{\mathrm{ir}}$ is independent of the
electromagnetic energy, a fact noticed by Bekenstein \cite{B71}. We have taken
one further step here by identifying the independent physical contributions to
$M_{\mathrm{ir}}$. This will have important consequences for the energetics of
black hole formation (see \cite{RV02}).

Next we consider the physical interpretation of the electromagnetic term
$\tfrac{Q^{2}}{2R}$, which can be obtained by evaluating the conserved Killing
integral
\begin{equation}
\int_{\Sigma_{t}^{+}}\xi_{+}^{\mu}T_{\mu\nu}^{\mathrm{(em)}}d\Sigma^{\nu}%
=\int_{R}^{\infty}r^{2}dr\int_{0}^{1}d\cos\theta\int_{0}^{2\pi}d\phi
\ T^{\mathrm{(em)}}{}{}_{0}{}^{0}=\tfrac{Q^{2}}{2R}\,, \label{EQR}%
\end{equation}
where $\Sigma_{t}^{+}$ is the space-like hypersurface in $\mathcal{M}_{+}$
described by the equation $t_{+}=t=\mathrm{const}$, with $d\Sigma^{\nu}$ as
its surface element vector and where $T_{\mu\nu}^{\mathrm{(em)}}=-\tfrac
{1}{4\pi}\left(  F_{\mu}{}^{\rho}F_{\rho\nu}+\tfrac{1}{4}g_{\mu\nu}%
F^{\rho\sigma}F_{\rho\sigma}\right)  $ is the energy-momentum tensor of the
electromagnetic field. The quantity in Eq.~(\ref{EQR}) differs from the purely
electromagnetic energy
\[
\int_{\Sigma_{t}^{+}}n_{+}^{\mu}T_{\mu\nu}^{\mathrm{(em)}}d\Sigma^{\nu}%
=\tfrac{1}{2}\int_{R}^{\infty}dr\sqrt{g_{rr}}\tfrac{Q^{2}}{r^{2}},
\]
where $n_{+}^{\mu}=f_{+}^{-1/2}\xi_{+}^{\mu}$ is the unit normal to the
integration hypersurface and $g_{rr}=f_{+}$. This is similar to the analogous
situation for the total energy of a static spherical star of energy density
$\epsilon$ within a radius $R$, $m\left(  R\right)  =4\pi\int_{0}^{R}%
dr\ r^{2}\epsilon$, which differs from the pure matter energy $m_{\mathrm{p}%
}\left(  R\right)  =4\pi\int_{0}^{R}dr\sqrt{g_{rr}}r^{2}\epsilon$ by the
gravitational energy (see \cite{MTW73}). Therefore the term $\tfrac{Q^{2}}%
{2R}$ in the mass formula (\ref{EQC}) is the \emph{total} energy of the
electromagnetic field and includes its own gravitational binding energy. This
energy is stored throughout the region $\mathcal{M}_{+}$, extending from $R$
to infinity.

\begin{figure}[ptb]
\begin{center}
\includegraphics[width=0.75\textwidth]{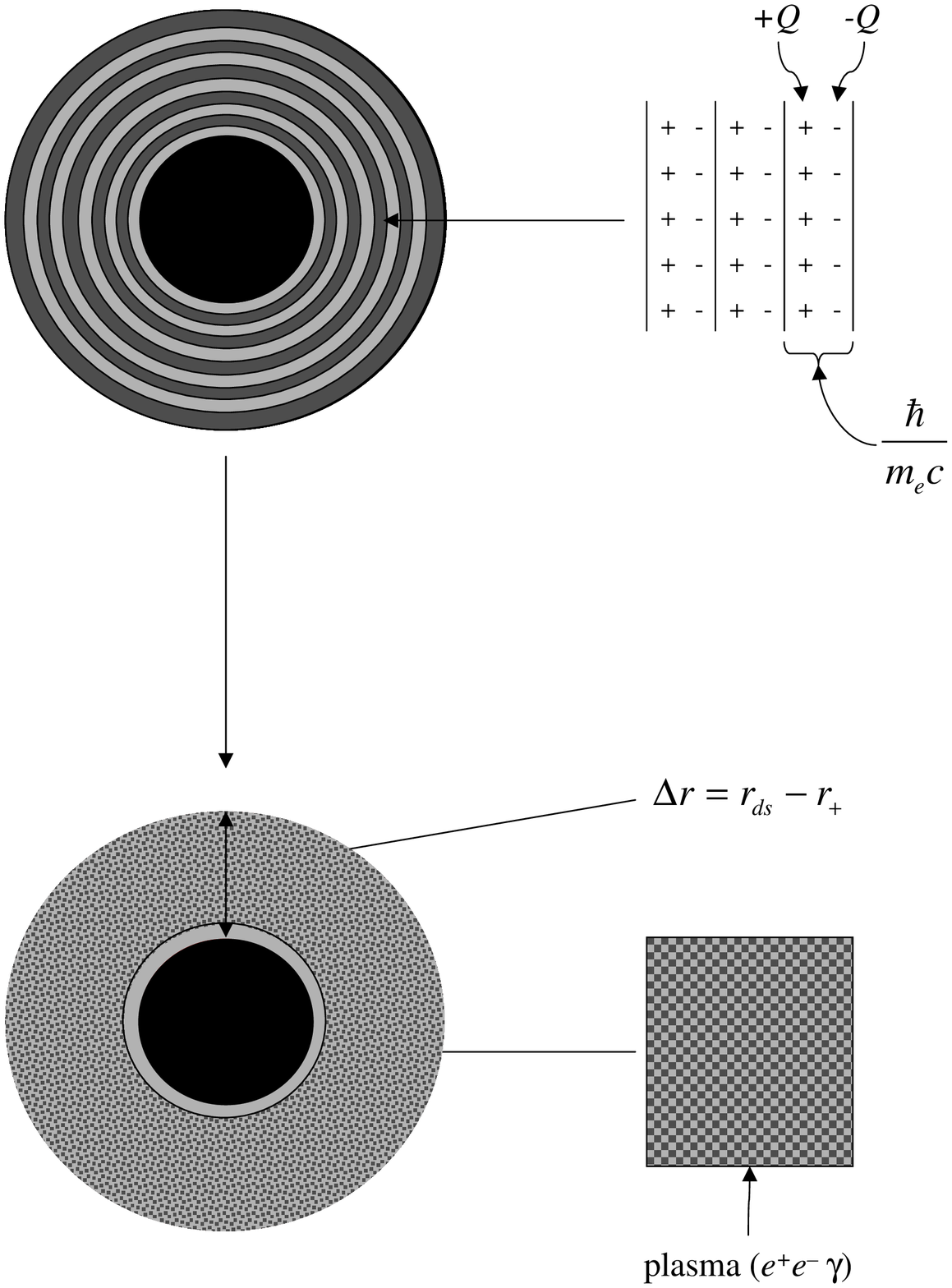}
\end{center}
\caption{Vacuum polarization process of energy extraction from an EMBH. Pairs
are created by vacuum polarization in the dyadosphere and the system
thermalizes to a neutral plama configuration (see \cite{PRX98,RBCFX01a} for
details).}%
\label{fig1}%
\end{figure}

We now turn to the problem of extracting the electromagnetic energy from an
EMBH (see \cite{CR71}). We can distinguish between two conceptually physically
different processes, depending on whether the electric field strength
$\mathcal{E}=\frac{Q}{r^{2}}$ is smaller or greater than the critical value
$\mathcal{E}_{\mathrm{c}}=\tfrac{m_{e}^{2}c^{3}}{e\hbar}$. Here $m_{e}$ and
$e$ are the mass and the charge of the electron. We recall that an electric
field $\mathcal{E}>\mathcal{E}_{\mathrm{c}}$ polarizes the vacuum creating
electron-positron pairs (see \cite{HE31,S51,DR75}). The maximum value
$\mathcal{E}_{+}=\tfrac{Q}{r_{+}^{2}}$ of the electric field around an EMBH is
reached at the horizon. We then have the following:

\begin{enumerate}
\item  For $\mathcal{E}_{+}<\mathcal{E}_{\mathrm{c}}$ the leading energy
extraction mechanism consists of a sequence of descrete elementary decay
processes of a particle into two oppositely charged particles. The condition
$\mathcal{E}_{+}<\mathcal{E}_{\mathrm{c}}$ implies
\begin{equation}
\xi\equiv\tfrac{Q}{\sqrt{G}M}\lesssim\left\{
\begin{array}
[c]{r}%
\tfrac{GM/c^{2}}{\lambda_{\mathrm{C}}}\left(  \tfrac{e}{\sqrt{G}m_{e}}\right)
^{-1}\sim10^{-6}\tfrac{M}{M_{\odot}}\quad\text{if }\tfrac{M}{M_{\odot}}%
\leq10^{6}\\
1\quad\quad\quad\quad\quad\quad\quad\quad\text{if }\tfrac{M}{M_{\odot}}>10^{6}%
\end{array}
\right.  ,\label{critical3}%
\end{equation}
where $\lambda_{\mathrm{C}}$ is the Compton wavelength of the electron.
Denardo and Ruffini \cite{DR73} and Denardo, Hively and Ruffini \cite{DHR74}
have defined as the \emph{effective ergosphere} the region around an EMBH
where the energy extraction processes occur. This region extends from the
horizon $r_{+}$ up to a radius
\begin{equation}
r_{\mathrm{Eerg}}=\tfrac{GM}{c^{2}}\left[  1+\sqrt{1-\xi^{2}\left(
1-\tfrac{e^{2}}{G{m_{e}^{2}}}\right)  }\right]  \simeq\tfrac{e}{m_{e}}%
\tfrac{Q}{c^{2}}\,.\label{EffErg}%
\end{equation}
The energy extraction occurs in a finite number $N_{\mathrm{PD}}$ of such
discrete elementary processes, each one correponding to a decrease of the EMBH
charge. We have
\begin{equation}
N_{\mathrm{PD}}\simeq\tfrac{Q}{e}\,.
\end{equation}
Since the total extracted energy is (see Eq.~(\ref{irrmass})) $E^{\mathrm{tot}%
}=\tfrac{Q^{2}}{2r_{+}}$, we obtain for the mean energy per accelerated
particle $\left\langle E\right\rangle _{\mathrm{PD}}=\tfrac{E^{\mathrm{tot}}%
}{N_{\mathrm{PD}}}$
\begin{equation}
\left\langle E\right\rangle _{\mathrm{PD}}=\tfrac{Qe}{2r_{+}}=\tfrac{1}%
{2}\tfrac{\xi}{1+\sqrt{1-\xi^{2}}}\tfrac{e}{\sqrt{G}m_{e}}\ m_{e}c^{2}%
\simeq\tfrac{1}{2}\xi\tfrac{e}{\sqrt{G}m_{e}}\ m_{e}c^{2},
\end{equation}
which gives
\begin{equation}
\left\langle E\right\rangle _{\mathrm{PD}}\lesssim\left\{
\begin{array}
[c]{r}%
\tfrac{M}{M_{\odot}}10^{21}eV\quad\text{if }\tfrac{M}{M_{\odot}}\leq10^{6}\\
10^{27}eV\quad\quad\text{if }\tfrac{M}{M_{\odot}}>10^{6}%
\end{array}
\right.  .\label{UHECR}%
\end{equation}
One of the crucial aspects of the energy extraction process from an EMBH is
its back reaction on the irreducible mass expressed in \cite{CR71}. Although
the energy extraction processes can occur in the entire effective ergosphere
defined by Eq. (\ref{EffErg}), only the limiting processes occurring on the
horizon with zero kinetic energy can reach the maximum efficiency while
approaching the condition of total reversibility (see Fig. 2 in \cite{CR71}
for details). The farther from the horizon that a decay occurs, the more it
increases the irreducible mass and loses efficiency. Only in the complete
reversibility limit \cite{CR71} can the energy extraction process from an
extreme EMBH reach the upper value of $50\%$ of the total EMBH energy.

\item  For $\mathcal{E}_{+}\geq\mathcal{E}_{\mathrm{c}}$ the leading
extraction process is a \emph{collective} process based on an electron-positron
plasma generated by the vacuum polarization, see Fig.~\ref{fig1}. The
condition $\mathcal{E}_{+}\geq\mathcal{E}_{\mathrm{c}}$ implies
\begin{equation}
\tfrac{GM/c^{2}}{\lambda_{\mathrm{C}}}\left(  \tfrac{e}{\sqrt{G}m_{e}}\right)
^{-1}\simeq2\cdot10^{-6}\tfrac{M}{M_{\odot}}\leq\xi\leq1\,.
\end{equation}
This vacuum polarization process can occur only for an EMBH with mass smaller
than $2\cdot10^{6}M_{\odot}$. The electron-positron pairs are now produced in
the dyadosphere of the EMBH, a subregion of the effective ergosphere which has
been defined in \cite{PRX98} and whose radius $r_{\mathrm{dya}}$ satisfies
$\mathcal{E}_{\mathrm{c}}\equiv\tfrac{Q}{r_{\mathrm{dya}}^{2}}.$ We have
\begin{equation}
r_{\mathrm{dya}}=\sqrt{\tfrac{eQ\hbar}{m_{e}^{2}c^{3}}}\,\ll r_{\mathrm{Eerg}%
}. \label{dya1}%
\end{equation}
The number of particles created \cite{PRX98} is then
\begin{equation}
N_{\mathrm{dya}}=\tfrac{1}{3}\left(  \tfrac{r_{\mathrm{dya}}}{\lambda
_{\mathrm{C}}}\right)  \left(  1-\tfrac{r_{+}}{r_{\mathrm{dya}}}\right)
\left[  4+\tfrac{r_{+}}{r_{\mathrm{dya}}}+\left(  \tfrac{r_{+}}%
{r_{\mathrm{dya}}}\right)  ^{2}\right]  \tfrac{Q}{e}\simeq\tfrac{4}{3}\left(
\tfrac{r_{\mathrm{dya}}}{\lambda_{\mathrm{C}}}\right)  \tfrac{Q}{e}\,.
\label{numdya}%
\end{equation}
The total energy stored in the dyadosphere is \cite{PRX98}
\begin{equation}
E_{\mathrm{dya}}^{\mathrm{tot}}=\left(  1-\tfrac{r_{+}}{r_{\mathrm{dya}}%
}\right)  \left[  1-\left(  \tfrac{r_{+}}{r_{\mathrm{dya}}}\right)
^{4}\right]  \tfrac{Q^{2}}{2r_{+}}\simeq\tfrac{Q^{2}}{2r_{+}}\,.
\label{enedya}%
\end{equation}
The mean energy per particle produced in the dyadosphere $\left\langle
E\right\rangle _{\mathrm{dya}}=\tfrac{E_{\mathrm{dya}}^{\mathrm{tot}}%
}{N_{\mathrm{dya}}}$ is then
\begin{equation}
\left\langle E\right\rangle _{\mathrm{dya}}=\tfrac{3}{2}\tfrac{1-\left(
\tfrac{r_{+}}{r_{\mathrm{dya}}}\right)  ^{4}}{4+\tfrac{r_{+}}{r_{\mathrm{dya}%
}}+\left(  \tfrac{r_{+}}{r_{\mathrm{dya}}}\right)  ^{2}}\left(  \tfrac
{\lambda_{\mathrm{C}}}{r_{\mathrm{dya}}}\right)  \tfrac{Qe}{r_{+}}\simeq
\tfrac{3}{8}\left(  \tfrac{\lambda_{\mathrm{C}}}{r_{\mathrm{dya}}}\right)
\tfrac{Qe}{r_{+}}\,, \label{meanenedya}%
\end{equation}
which can be also rewritten as
\begin{equation}
\left\langle E\right\rangle _{\mathrm{dya}}\simeq\tfrac{3}{8}\left(
\tfrac{r_{\mathrm{dya}}}{r_{+}}\right)  \ m_{e}c^{2}\sim\sqrt{\tfrac{\xi
}{M/M_{\odot}}}10^{5}keV\,. \label{GRB}%
\end{equation}
Such a process of vacuum polarization around an EMBH has been observed to
reach the maximum efficiency limit of $50\%$ of the total mass-energy of an
extreme EMBH (see e.g. \cite{PRX98}). The conceptual justification of this
result needs, however, the dynamical analysis of the vacuum polarization
process during the gravitational collapse and the implementation of the
screening of the $e^{+}e^{-}$ neutral plasma generated in this process. This
analysis based on our present work conceptually validates the reversibility of
the process and is given in \cite{RVX02}.
\end{enumerate}

Let us now compare and contrast these two processes. We have
\begin{equation}
r_{\mathrm{Eerg}}\simeq\left(  \tfrac{r_{\mathrm{dya}}}{\lambda_{\mathrm{C}}%
}\right)  r_{\mathrm{dya}},\quad N_{\mathrm{dya}}\simeq\left(  \tfrac
{r_{\mathrm{dya}}}{\lambda_{\mathrm{C}}}\right)  N_{\mathrm{PD}},\quad
\left\langle E\right\rangle _{\mathrm{dya}}\simeq\left(  \tfrac{\lambda
_{\mathrm{C}}}{r_{\mathrm{dya}}}\right)  \left\langle E\right\rangle
_{\mathrm{PD}}.
\end{equation}
Moreover we see (Eqs. (\ref{UHECR}), (\ref{GRB})) that $\left\langle
E\right\rangle _{\mathrm{PD}}$ is in the range of energies of UHECR (see
\cite{NW00} and references therein), while for $\xi\sim0.1$ and $M\sim
10M_{\odot}$, $\left\langle E\right\rangle _{\mathrm{dya}}$ is in the gamma
ray range. In other words, the discrete particle decay process involves a
small number of particles with ultra high energies ($\sim10^{21}eV$), while
vacuum polarization involves a much larger number of particles with lower mean
energies ($\sim10MeV$).

The new conceptual understanding of the mass formula presented here has
important consequences for the energetics of a black hole. The expression for
the irreducible mass in terms of its different physical constituents (Eq.
(\ref{EQM})) leads to a reinterpretation of the energy extraction process
during the formation of a black hole as expressed in \cite{RV02}. It will
certainly be interesting to reach an understanding of the new expression for
the irreducible mass in terms of its thermodynamical analogues.

The energy extraction processes from an EMBH are shown here to be separated
into two very different classes depending on the strength of the
electromagnetic field ($\mathcal{E}\lessgtr\mathcal{E}_{\mathrm{c}}$). The
process occurring for $\mathcal{E}<\mathcal{E}_{\mathrm{c}}$ leads to a
prolonged ($\tau\sim10^{2}-10^{4}yrs$) very high energy emission $E\sim
10^{21}eV$ (see \cite{PRX98}). This process can be the basis for an
explanation of UHECR \cite{CR02}. On the other hand it is clear to us now that
the process of vacuum polarization, whose key formulas are summarized in Eqs.
(\ref{dya1}), (\ref{numdya}), (\ref{enedya}), (\ref{meanenedya}), appears more
and more to be at the very heart of the solution of thirty years of
problematics in modeling GRBs \cite{RBCFX01a}.

\newpage

\end{document}